\documentclass[5p, sort&compress]{elsarticle}

\usepackage[utf8x]{inputenc}
\usepackage[T1]{fontenc}

\usepackage{microtype}

\usepackage{graphicx}
\usepackage{tabularx}
\usepackage{booktabs}
\usepackage{float}
\usepackage{siunitx}
\usepackage[version=4]{mhchem}
\usepackage[export]{adjustbox}[2011/08/13] %
\usepackage{pdfpages}
\usepackage{tikz}

\usepackage{hyperref} %

\newcommand{\mytitle}{%
  From metallic glasses to nanocrystals: Molecular dynamics
  simulations on the crossover from glass-like to
  grain-boundary-mediated deformation behaviour%
}

\hypersetup{
  final,
  pdfborder={0 0 0},
  pdftitle={\mytitle},
  pdfauthor={Tobias Brink and Karsten Albe},
  colorlinks=true,
  urlcolor=red,
  linkcolor=black,
  citecolor=blue
}

\begin{document}
\begin{frontmatter}
\title{\mytitle}

\author{Tobias Brink\corref{cor1}\fnref{epfl}}
\ead{brink@mm.tu-darmstadt.de}

\author{Karsten Albe\corref{cor2}} %

\cortext[cor1]{Corresponding author}
\address{Fachgebiet Materialmodellierung, Institut f{\"u}r
  Materialwissenschaft,\\Technische Universit\"at Darmstadt,
  Otto-Berndt-Stra\ss{}e~3, D-64287 Darmstadt, Germany}
\fntext[epfl]{Present address: \textit{Civil Engineering Institute and
                Institute of Materials Science and Engineering,
                \'Ecole polytechnique f\'ed\'erale de Lausanne (EPFL),
                Station 18, CH-1015 Lausanne, Switzerland}}

\begin{abstract}
  Nanocrystalline metals contain a large fraction of high-energy grain
  boundaries, which may be considered as glassy phases.  Consequently,
  with decreasing grain size, a crossover in the deformation behaviour
  of nanocrystals to that of metallic glasses has been proposed.
  Here, we study this crossover using molecular dynamics simulations
  on bulk glasses, glass--crystal nanocomposites, and nanocrystals of
  \ce{Cu64Zr36} with varying crystalline volume fractions induced by
  long-time thermal annealing. We find that the grain boundary phase
  behaves like a metallic glass under constraint from the abutting
  crystallites. The transition from glass-like to
  grain-boundary-mediated plasticity can be classified into three
  regimes: (1) For low crystalline volume fractions, the system
  resembles a glass--crystal composite and plastic flow is localised
  in the amorphous phase; (2) with increasing crystalline volume
  fraction, clusters of crystallites become jammed and the mechanical
  response depends critically on the relaxation state of the glassy
  grain boundaries; (3) at grain sizes $\geq \SI{10}{nm}$, the system
  is jammed completely, prohibiting pure grain-boundary plasticity and
  instead leading to co-deformation. We observe an inverse Hall--Petch
  effect only in the second regime when the grain boundary is not
  deeply relaxed.  \raisebox{0pt}[\height][0pt]{%
    Experimental results with different grain boundary states are
    therefore not directly comparable in this regime.}
\end{abstract}


\journal{Acta Materialia, online at
  \href{https://doi.org/10.1016/j.actamat.2018.06.036}
       {\texttt{https:/\!/doi.org/10.1016/j.actamat.2018.06.036}}.}

\end{frontmatter}

\begin{tikzpicture}[remember picture,overlay]
  \node [anchor=north west, font=\footnotesize\itshape, align=left,
         xshift=-\oddsidemargin, yshift=1in]
        at (current page.south west)
        {\phantom{Pp}\\
         © 2018. This manuscript version is made available under the
         CC-BY-NC-ND 4.0 licence.\\
         \textup{\url{http://creativecommons.org/licenses/by-nc-nd/4.0/}}};
\end{tikzpicture}

\frenchspacing

\section{Introduction}

\begin{figure}[b!]
  \centering
  \includegraphics[center]{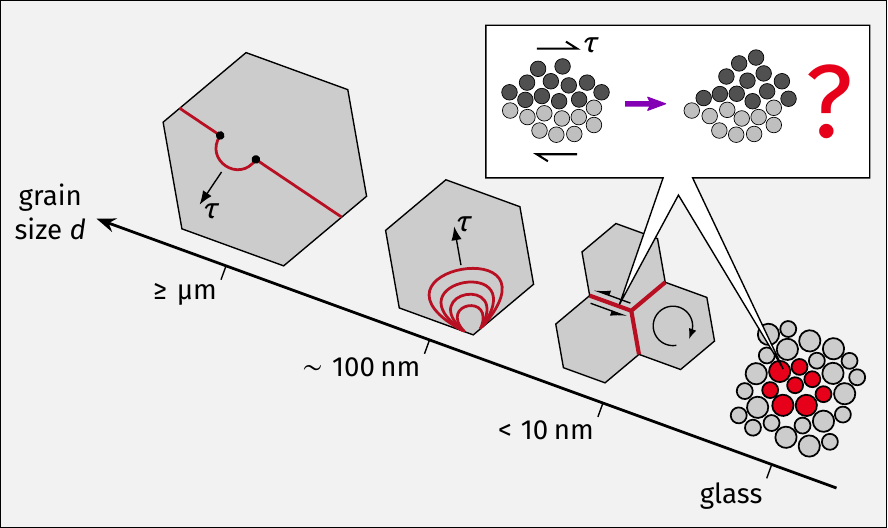}
  \caption{Transition from nanocrystals to metallic glasses. While
    coarse-grained polycrystals exhibit classical, bulk-like
    dislocation activity, deformation mechanisms in nanocrystals depend on the average grain size. Grain boundaries act as sinks and
    sources for dislocation activity, and the plasticity of the grain
    boundaries themselves becomes important in the sub-10-nm
    regime. With further decrease of the grain size $d$, a transition
    to deformation behaviour known from glasses was postulated:
    Experimental data indicate similarities between shear
    transformations in glasses and grain boundary plasticity,
    suggesting a smooth transition to the glassy state at small $d$.}
  \label{fig:intro}
\end{figure}

Deformation mechanisms operating in nanocrystalline metals sensitively depend on the average grain size and
grain size distribution \cite{Lohmiller2014}. In contrast to coarse-grained polycrystalline
metals, where plastic strain is predominantly carried by dislocations, grain-boundary-mediated deformation processes (including sliding and shuffling mechanisms) become increasingly relevant with decreasing grain size. Since the first controlled synthesis of nanocrystalline metals in the late 1980s \cite{Gleiter1989}, a large body of literature on the nature of deformation mechanisms in this material class has appeared, which eventually led to the development of deformation maps showing active mechanisms as a function of grain size, temperature, or strain rate (see Ref.~\citenum{Meyers2006} for an overview).
Since low-energy grain boundaries are typically absent in nanocrystalline microstructures \cite{Meyers2006}, several studies have fostered the view that a nanocrystalline metal could also be considered as a composite of a glassy, percolating grain boundary phase confined between nanocrystallites \cite{Wolf2005, Trelewicz2007}. Thus, a smooth transition to glass-like deformation behaviour could be expected with decreasing grain size (see Fig.~\ref{fig:intro}).
The parallels between grain boundary deformation and
the deformation mechanisms in metallic glasses (MGs) are indeed striking: MGs
deform via shear-transformation zones (STZs), small regions in which a
shear deformation is activated under external stress \cite{Argon1979,
  Falk1998}.  Similar concepts have been applied to grain boundaries
\cite{Lund2004, Argon2006} and yielding in nanocrystals has been shown
to exhibit a $T^{2/3}$ temperature dependence \cite{Grewer2014}, just
as described for MGs by Johnson and Samwer \cite{Johnson2005}.
Trelewicz and Schuh performed nanoindentation experiments on very
small-grained Ni--W alloys and found that they show pop-in events
similar to those observed in indentation on metallic glasses
\cite{Trelewicz2007}. Moreover, the indents featured shear offsets in the
surrounding pile-up, indicative of shear banding.
An extrapolation of these indentation testing data with varying grain sizes arrives at the amorphous limit \cite{Schuh2003, Detor2007, Trelewicz2007}, estimated at around \SI{1}{nm} and supported by simulations that report a collapse of the crystalline
lattice at such small grain sizes \cite{Wolf1995, Zheng2007, Brink2015}.
It was also conclusively shown by experiment and computer simulations that the grain boundary
phase is characterised by a lower shear modulus, which starts
affecting the effective macroscopic shear modulus \cite{Zhao2006, Grewer2011},
making the structure reminiscent of the liquid-like/solid-like
division in metallic glasses \cite{Dmowski2010a, Ding2012, Ding2013,
  Ding2014b}.

A problem with the experimental data, which was mostly obtained for
Ni--W systems, is that the grain size can only be controlled by
varying the tungsten content \cite{Schuh2003, Detor2007,
  Trelewicz2007}.  Pure systems exhibit rapid grain growth even at
room temperature \cite{Weissmueller1995, Ames2008}.  By segregation of
solutes to the grain boundaries, small grain sizes can be stabilised
\cite{Weissmueller1993, Kirchheim2002, Liu2004, Beke2004, Krill2005,
  Detor2007, Schaefer2012a}, even in miscible systems
\cite{Kurmanaeva2010, Schaefer2011}.  As a result, though, not only
the grain size but also the nature of the grain boundary itself may be
modified \cite{Udler1992, Rupert2011, Schaefer2012, Schaefer2012a,
  Schaefer2012b, Oezerinc2012}.  Therefore, any quantitative trend extracted
experimentally from varying grain sizes below \SI{10}{nm} has to be interpreted with care, since the grain boundary state and composition also changes with grain size.

If we accept at this point the proposition of glass-like grain
boundaries, we could also consider a nanocrystal as one limit of a
glass--crystal nanocomposite. Such composites have received a lot of
attention for their potential to retain the high yield strength and
large elastic limit of MGs, while improving their ductility. It is
generally found that secondary phases enhance the tendency for shear band
nucleation and lead to a more homogeneous strain distribution
\cite{Hajlaoui2007, Hofmann2008, Albe2013, Zaheri2014}. Regarding the
interaction of propagating shear bands with precipitates, small
precipitates can be avoided (``wrapped'') by shear bands
\cite{Brink2016}, leaving the crystalline phase undeformed
\cite{Pauly2010a}. Bigger precipitates block the shear band
propagation or co-deform with the glass matrix \cite{Pauly2010a,
  Brink2016}. These investigations are mostly concerned with smaller
volume fractions of crystalline phase, though, and it is unclear if
they also apply to nanocrystalline metals.

The purpose of the present work is to study the crossover from
glass--crystal nanocomposites to nanocrystals using molecular dynamics
(MD) computer simulations.  We use the well-established MG
\ce{Cu64Zr36} with embedded brittle Laves phase nanocrystallites as a
model system. The advantage over studying ductile nanocrystallites is
that the plastic response of the system is exclusively carried by the
glassy phase/grain boundaries and thus we can disentangle the grain
boundary activity from dislocation activity.  As was recently shown
\cite{Tang2012, Zemp2014, Zemp2015}, Laves crystallites can be grown
in MD simulations using the Finnis--Sinclair-type potential by
Mendelev \cite{Mendelev2009}. This opens the possibility of growing
crystallites in a system of constant composition, thereby obtaining
samples ranging from a homogeneous MG, over glass--crystal composites,
to nanocrystals.

\section{Computational methods and analysis}

All MD simulations were performed with Cu--Zr systems using
\textsc{lammps} \cite{Plimpton1995} and the Finnis--Sinclair-type
potential by Mendelev \textit{et al.} \cite{Mendelev2009}. The
integration time step was \SI{2}{fs} in all cases.

\subsection{Synthesis and annealing procedure}
\begin{figure}[b!]
  \centering
  \includegraphics[center]{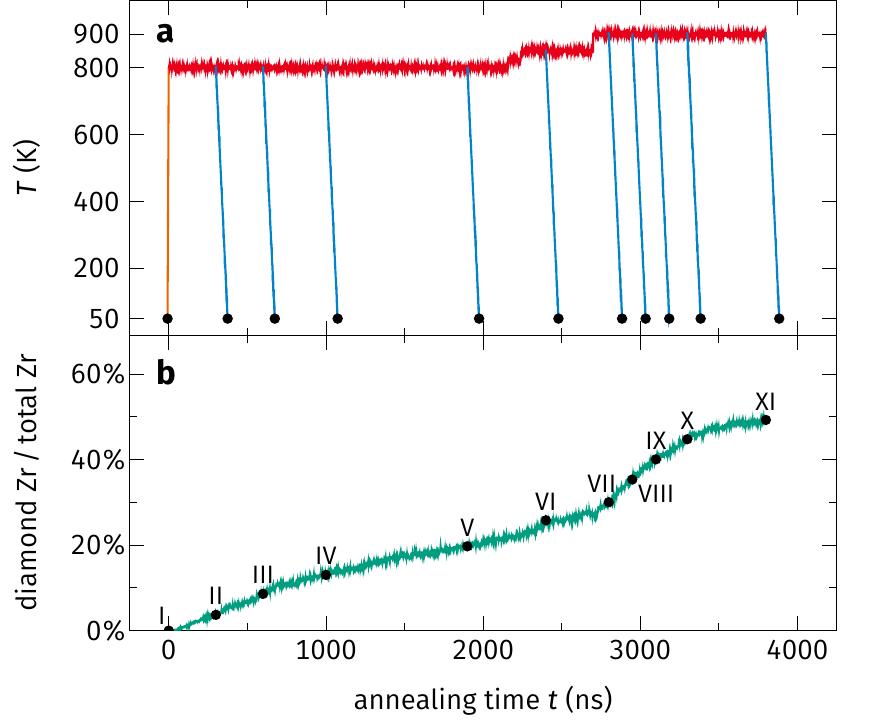}
  \caption{The annealing procedure. (a) Temperature profile. The
    sample is heated with \SI{0.1}{K/ps} to \SI{800}{K}. After around
    \SI{2}{\micro\second} the temperature is increased step-wise to
    increase the crystallisation rate. Snapshots are taken at
    intervals and cooled down with \SI{0.01}{K/ps}. (b) The fraction
    of zirconium atoms in diamond superlattice configurations as a function of
    annealing time as a rough indicator of crystalline volume
    fraction. The snapshots are labelled I--XI.}
  \label{fig:annealing-procedure}
\end{figure}

A \ce{Cu64Zr36} metallic glass consisting of $N = 63,108$ atoms was
quenched from the melt down to \SI{50}{K} with a cooling rate of
\SI{0.01}{K/ps} and equilibrated there for \SI{2}{ns}. This
``as-quenched'' sample is designated sample~I and has dimensions of
around \SI{10x10x10}{nm}.  In order to facilitate the crystallisation
of \ce{Cu2Zr} Laves phases \cite{Zemp2015}, we heated the glass to
\SI{800}{K} (which is close to the glass transition $T_g$
\cite{Mendelev2009}) with $\dot{T} = \SI{0.1}{K/ps}$. We held the
glass at this temperature for roughly \SI{2}{\micro\second} and then
increased the temperature step-wise to \SI{850}{K} and then to
\SI{900}{K} to speed up the crystallisation kinetics.
Figure~\ref{fig:annealing-procedure} shows the complete annealing
procedure. Neither a glass transition nor melting were observed;
indeed the glass transition temperature rises with annealing time
(see \ref{sec:appendix:Tg}).  At intermediate steps, snapshots of the
simulation were taken and cooled with $\dot{T} = \SI{0.01}{K/ps}$ back
to \SI{50}{K}.  These snapshots with different crystalline volume
fractions are labelled as sample~II--XI. We detected the presence of
crystallites and computed the grain sizes as described in
\ref{sec:appendix:crystanal} and found a mix of C14 and C15 Laves
phases \cite{DeGraef2007} with an amorphous grain boundary.  Details
about the nucleation of crystallites are provided in
\ref{sec:appendix:nucleation}.

For later mechanical testing, these samples were replicated
$3 \times 1 \times 7$ times to obtain specimen of \SI{30x10x70}{nm}
size.  Open boundaries were introduced in $x$ direction to increase
the tendency for shear localisation as discussed in
Ref.~\citenum{Albe2013}.  The samples were then equilibrated at 
target temperatures of \SI{50}{K} and \SI{250}{K} for \SI{1}{ns}.  The
structural, thermodynamic, and elastic properties were determined
using the unreplicated samples with periodic boundaries.

\subsection{Artificial nanocrystals by Voronoi construction}

As a comparison, we also prepared artificial nanocrystals by choosing
a number of random points in a simulation cell, constructing Voronoi
cells \cite{Voronoi1908a} around those points, and inserting either a
C14 or a C15 lattice with random orientation into the cells. This
procedure is analogous to the one described in
Ref.~\citenum{Derlet2003}.  Assuming spherically shaped grains, we can
obtain a given grain diameter $d$ in a simulation cell of volume $V$
using $n = \left\lfloor 6V / (\pi d^3) \right\rfloor$ points.  We
selected approximate grain sizes of \SI{3}{nm}, corresponding to the
grain size in sample XI, as well as \SI{5}{nm}, \SI{7}{nm},
\SI{10}{nm}, and \SI{15}{nm}, which are closer to the grain sizes
obtainable in experiment.  The samples were prepared with dimensions
similar to the grown crystallite composites and also have open
boundaries in $x$ direction.  We optimised the resulting structures by
molecular statics simulations and equilibrated them at the target
temperatures of \SI{50}{K} and \SI{250}{K} for \SI{1}{ns}.

\subsection{Relaxation state of the amorphous phase}

Metallic glasses exhibit a direct correlation between yield stress and
shear modulus $G$ \cite{Johnson2005}, which means that the shear modulus
can also serve as an indicator for the relaxation state of the glass.
However, the determination of local moduli for grain boundaries is
nontrivial, since (1) one needs to measure a localised shear modulus
and (2) amorphous metals are anisotropic on the nanometre length scale
\cite{Luo2015}.  We therefore use the Kelvin notation for the
stiffness tensor \cite{Thomson1856}, which we calculate per atom as
described in Ref.~\citenum{Brink2016a}. Diagonalisation of this tensor
yields five shear moduli \cite{Derlet2012}. We use the median of the
smallest of these, $\langle G_1\rangle$, as an indicator for the
overall shear stiffness of the amorphous phase, assuming that the
``softest'' shear mode is the one that yields first.  More details on
the procedure and numerical concerns are discussed in
\ref{sec:appendix:elprop}.

\subsection{Tensile tests}

For mechanical testing a constant engineering strain rate of
\SI{e8}{s^{-1}} was applied uniaxially in $z$ direction of the
equilibrated samples up to a maximum strain of $12\%$, which exceeds
the yield strain. The deformation was performed for each sample at
\SI{50}{K} and \SI{250}{K}.
Localised deformation was identified using the atomic shear strain
$\eta_i$ \cite{Shimizu2007} as implemented in \textsc{ovito}
\cite{Stukowski2010}.  As a measure for the degree of localisation we
use the shear localisation parameter
$\psi = \sqrt{\sum_{i=1}^N \left(\eta_i - \overline{\eta}\right)^2 /
  N}$, with $\overline{\eta} = \sum_{i=1}^N \eta_i / N$, where $N$ is
the number of atoms in the system \cite{Cheng2009}.

\begin{figure}[b!]
  \centering
  \includegraphics[center]{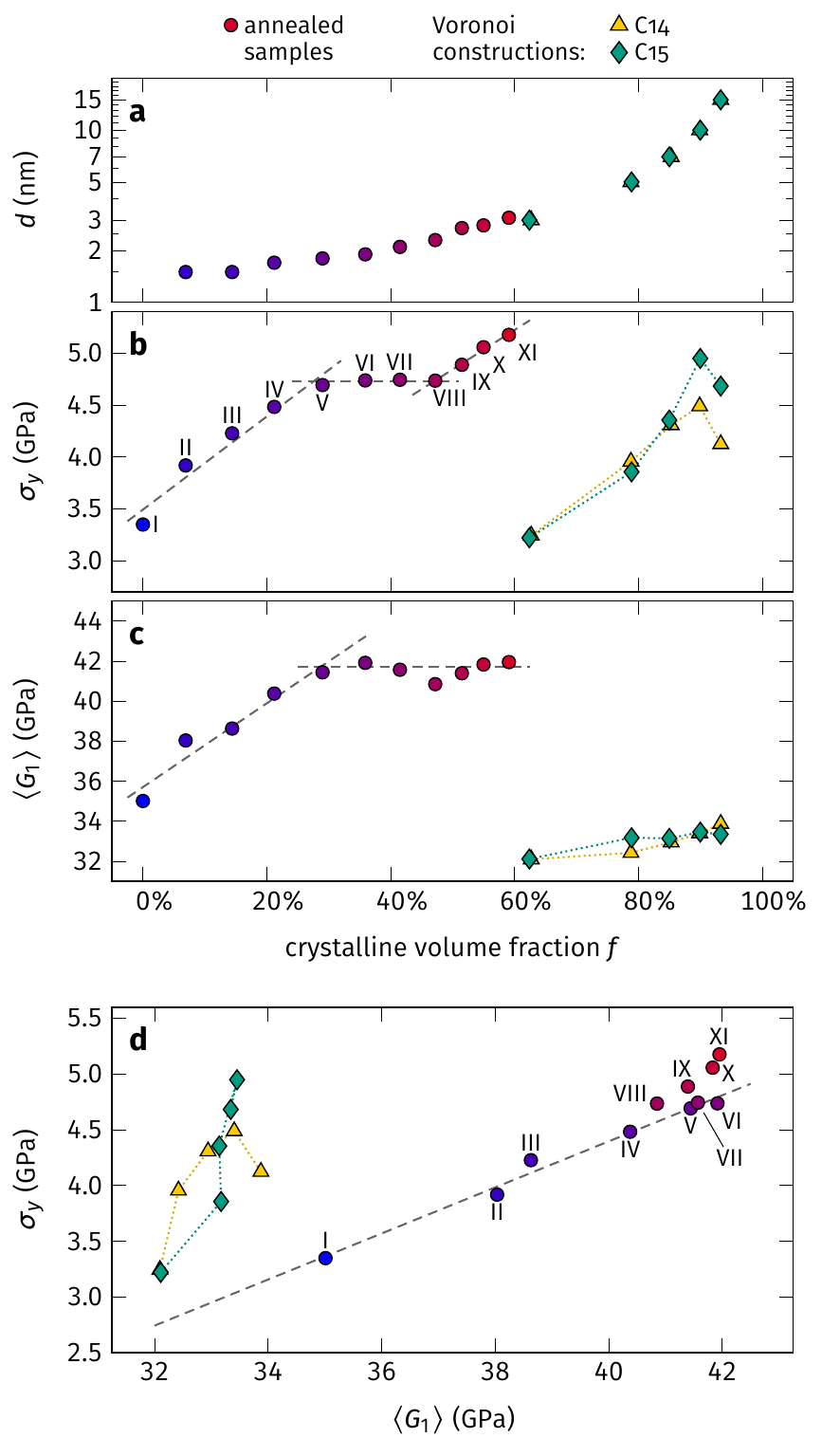}
  \caption{Characterisation of the samples and their mechanical
    strength. (a) Microstructure as defined by average crystallite
    diameter $d$ and crystalline volume fraction $f$. Yield stress at
    \SI{50}{K} (b) and median atomic shear modulus
    $\langle G_1 \rangle$ of the amorphous phase (c) reveal a
    proportionality $\sigma_y \propto \langle G_1 \rangle$ that breaks
    down at higher $f$. This is again illustrated in (d) by plotting
    both values against each other. The dashed lines serve as a guide
    for the eye: Samples VIII--XI no longer exhibit the
    proportionality; neither do the samples created artificially by
    Voronoi construction. A direct relation to $d$ cannot be observed.}
  \label{fig:yield-over-shearmod}
\end{figure}
\begin{figure*}
  \centering
  \includegraphics[center]
                  {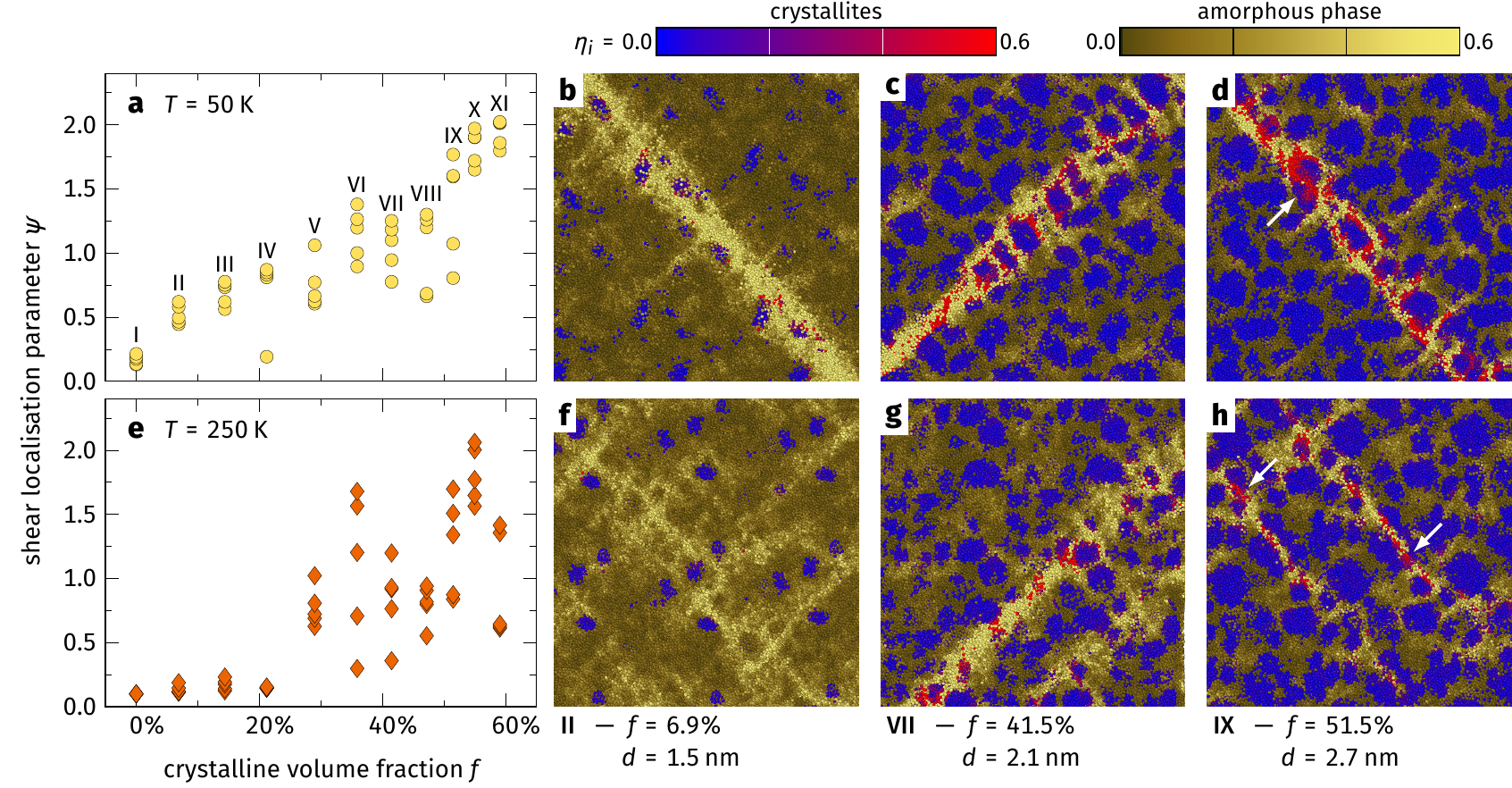}
  \caption{Shear localisation at $\varepsilon=12\%$ in samples
    I--XI. The top row shows the results for $T = \SI{50}{K}$, the
    bottom row for $T = \SI{250}{K}$. (a) and (e) show the shear
    localisation parameter $\psi$. The snapshots in (b)--(d) and
    (f)--(h) are coloured according to the atomic shear strain $\eta_i$
    using two different colour maps for the amorphous and the
    crystalline phase.  Crystallites that are cut are marked by
    arrows. More snapshots are provided in Supplementary Fig.~S.5 and S.6.}
  \label{fig:shear-localization-annealed}
\end{figure*}

We define the yield stress $\sigma_y$ of the samples as the maximum
stress in the stress--strain curve. We use this criterion instead of
the flow stress, since all phases in the system will soften when damaged.
The crystallites are brittle and thus are destroyed when yielding, and the
glass undergoes shear softening. Indeed, differently relaxed glasses of
the same composition exhibit different maximum stresses but the same flow
stress \cite{Cheng2008, Cheng2009}.

\section{Results}
\subsection{Mechanical properties and deformation mechanisms}

Characterisation of the annealed samples (I--XI) shows that the
average crystallite diameters $d$ are quite small ($\leq \SI{3}{nm}$)
and therefore suitable to explore the sub-10-nm nanocrystalline regime
(Fig.~\ref{fig:yield-over-shearmod}a).
The results of tensile tests at \SI{50}{K} on these specimen are shown
in Fig.~\ref{fig:yield-over-shearmod}b. Tests on samples I--XI were
repeated five times, but the variation of the yield stress is
negligible and the error bars are smaller than the symbols.  The
results at \SI{250}{K} are qualitatively similar apart from a uniform
decrease of yield strength and are omitted here. They are provided in
the Supplementary Fig.~S.1. Full stress--strain curves are available
in Supplementary Fig.~S.2. The general trend is an increase of yield
strength $\sigma_y$ with crystalline volume fraction $f$. All samples
deform by shear banding, except for samples I and II, which exhibit a
somewhat more homogeneous deformation at \SI{250}{K} (see
Fig.~\ref{fig:shear-localization-annealed} and Supplementary Figs.~S.3
and S.4). We note here that nucleation
phenomena at the crystal--glass interface seem to play a minor role
for the strength of the composite, otherwise the differences in yield
stresses between \SI{50}{K} and \SI{250}{K} would be more complex than
a simple offset and depend on the interface area.

According to earlier simulations \cite{Brink2016},
crystallites with diameters $d$ of around \SI{3}{nm} should pose no
obstacle to shear band propagation and the yield stress should be
defined by the amorphous phase alone.  The yield stress of
the amorphous phase is proportional to its shear modulus $G$
\cite{Johnson2005}.  Figures~\ref{fig:yield-over-shearmod}b,
\ref{fig:yield-over-shearmod}c, and \ref{fig:yield-over-shearmod}d
show that the proportionality between the local shear modulus $\langle
G_1\rangle$ in the grain boundary and the yield stress
$\sigma_y$ works for low volume fractions $f$: During the annealing
process, the amorphous phase relaxes and becomes stiffer.  For large
volume fractions of the crystalline phase, though, the proportionality
with $\langle G_1\rangle$ breaks down: The glass reaches a maximum
stiffness (and therefore a deeply relaxed state), while the yield
stress increases further from sample VIII on.  At this point the
microstructure starts to play a role, since only the crystalline
volume fraction $f$ and grain size $d$ continue changing.  This also
marks the breakdown of the typical mechanisms in glass--crystal
composites, where shear bands can always avoid such small crystallites
\cite{Brink2016}.

\begin{figure*}
  \centering
  \includegraphics[center]
                  {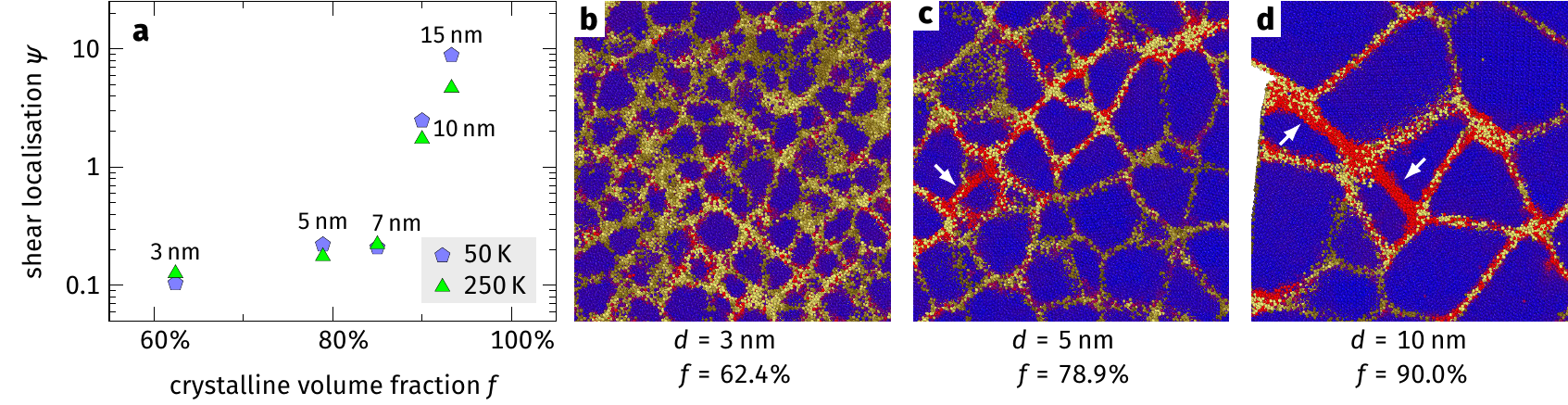}
  \caption{Shear localisation at $\varepsilon=12\%$ in the samples
    produced by Voronoi construction. (a) The shear localisation
    parameter $\psi$ shows that the deformation is rather homogeneous
    independent of the temperature, indicating that the grain boundary
    is comparatively weak and plastic events do not require thermal
    activation. (b)--(d) Snapshots of the atomic shear strain after deformation at \SI{50}{K}. The colour scale is the same as in
    Fig.~\ref{fig:shear-localization-annealed}. More snapshots are
    provided in Supplementary Figs.~S.5 and S.6.}
  \label{fig:shear-localization-voronoi}
\end{figure*}
As a first step towards understanding this transition, we evaluated
the strain localisation behaviour of the samples.
Figure~\ref{fig:shear-localization-annealed} and Supplementary
Figs.~S.3 and S.4 show that the shear localisation increases together
with the crystalline volume fraction and decreases with temperature.
Clearly, STZ activation is a thermally activated process
\cite{Argon1979} and thus an increased temperature leads to increased
activation kinetics all over the sample and a more homogeneous
deformation. The change of strain localisation with $f$ can be
explained by competing mechanisms: The reduction of the amorphous
volume fraction leads to a spatial confinement of STZ activation and
necessarily increases the heterogeneity of the early deformation
stages. This is exacerbated by the increasing heterogeneity of the
glassy phase itself: Glasses that are deeply relaxed are defined by a
greater difference between liquid-like and solid-like regions, making
the deformation once again more heterogeneous \cite{Cheng2009}. On the
other hand, dispersions of small crystallites in a glass matrix were
observed to increase the density of shear bands
\cite{Hajlaoui2007}. Therefore, the shear localisation stays low for
the samples I--IV at \SI{250}{K}. Here, the increased temperature
enables nucleation of multiple shear bands which is supported by the
dispersion of crystallites which serve as shear band nucleation sites
\cite{Albe2013, Zaheri2014, Wang2014b}. Starting from sample V, the
shear bands become more confined and the further growth of the nascent
shear bands into mature ones is suppressed.  It can be seen in
Fig.~\ref{fig:shear-localization-annealed}, as well as in
Supplementary Figs.~S.3 and S.4, that patterns of localised
deformation persist in samples with high $f$, indicating that
shear band multiplication still occurs, but that the growth of a
multitude of shear bands is suppressed. Consequently, strain
localisation is increased.  Furthermore,
Figs.~\ref{fig:shear-localization-annealed}d and
\ref{fig:shear-localization-annealed}h show that the confinement
additionally leads to the crystallites being cut.  Since the
crystallites are brittle and stronger than the glass, the yield stress
increases, even if the glass phase itself does not harden.  Due to the
brittleness of the Laves phases, the cut occurs in the form of
amorphisation along a favourable lattice plane.

In order to sample more geometries, we also artificially created
nanocrystals with \SI{3}{nm} to \SI{15}{nm} grain sizes using the
Voronoi construction algorithm \cite{Derlet2003}.  We constructed the
samples once with the C14 and once with the C15 Laves phase
\cite{DeGraef2007}, which both occur during the annealing process.  We
performed the same tensile tests and analyses as for samples I--XI and
show the results in Fig.~\ref{fig:yield-over-shearmod}. Interestingly,
the artificial sample with \SI{3}{nm} grain size has a drastically
reduced strength compared with sample XI, despite their similar
microstructure. Given the previously discussed results, one would
expect that the crystallites partake in the deformation and that both
samples exhibit the same yield stress.  This is not the case.
Figure~\ref{fig:shear-localization-voronoi}b shows that the
deformation in the amorphous phase is delocalised and that this
homogeneous flow can occur without involvement of the crystalline
phase.  The reason can be found in
Fig.~\ref{fig:yield-over-shearmod}c: The grain boundary of the
artificially constructed samples is very soft, indicating a
high-energy state of the amorphous phase.  It is known that this leads
to delocalised deformation and lower yield stress \cite{Cheng2009}.
Figure~\ref{fig:shear-localization-voronoi}b confirms that the grain
boundary phase ``flows around'' the crystallites.  With increasing
grain size and crystalline volume fraction, though, more and more
crystallites are being cut (see
Figs.~\ref{fig:shear-localization-voronoi}c and
\ref{fig:shear-localization-voronoi}d), despite comparable softness
of the grain boundary.  This is accompanied by an increase in strain
localisation (Fig.~\ref{fig:shear-localization-voronoi}a).  As a
result, an inverse Hall--Petch relation (cf.\
Refs.~\citenum{Schuh2003, VanVliet2003, Trelewicz2007}) occurs. The
hardening with increasing grain size can be explained by the
increasing fraction of crystalline matter that has to deform plastically.
Starting with \SI{10}{nm} grain size, the typical order of magnitude
for the breakdown of the Hall--Petch relation \cite{Schuh2003,
  Detor2007, Trelewicz2007}, shear localisation becomes very high and
the yield stress reaches a plateau comparable in magnitude to sample
XI.  This signifies the transition from a macroscopic plastic response
dominated by the grain boundary to a response dominated mainly
(although not exclusively) by the grain interior.

\begin{figure*}[t!]
  \centering
  \includegraphics[center]{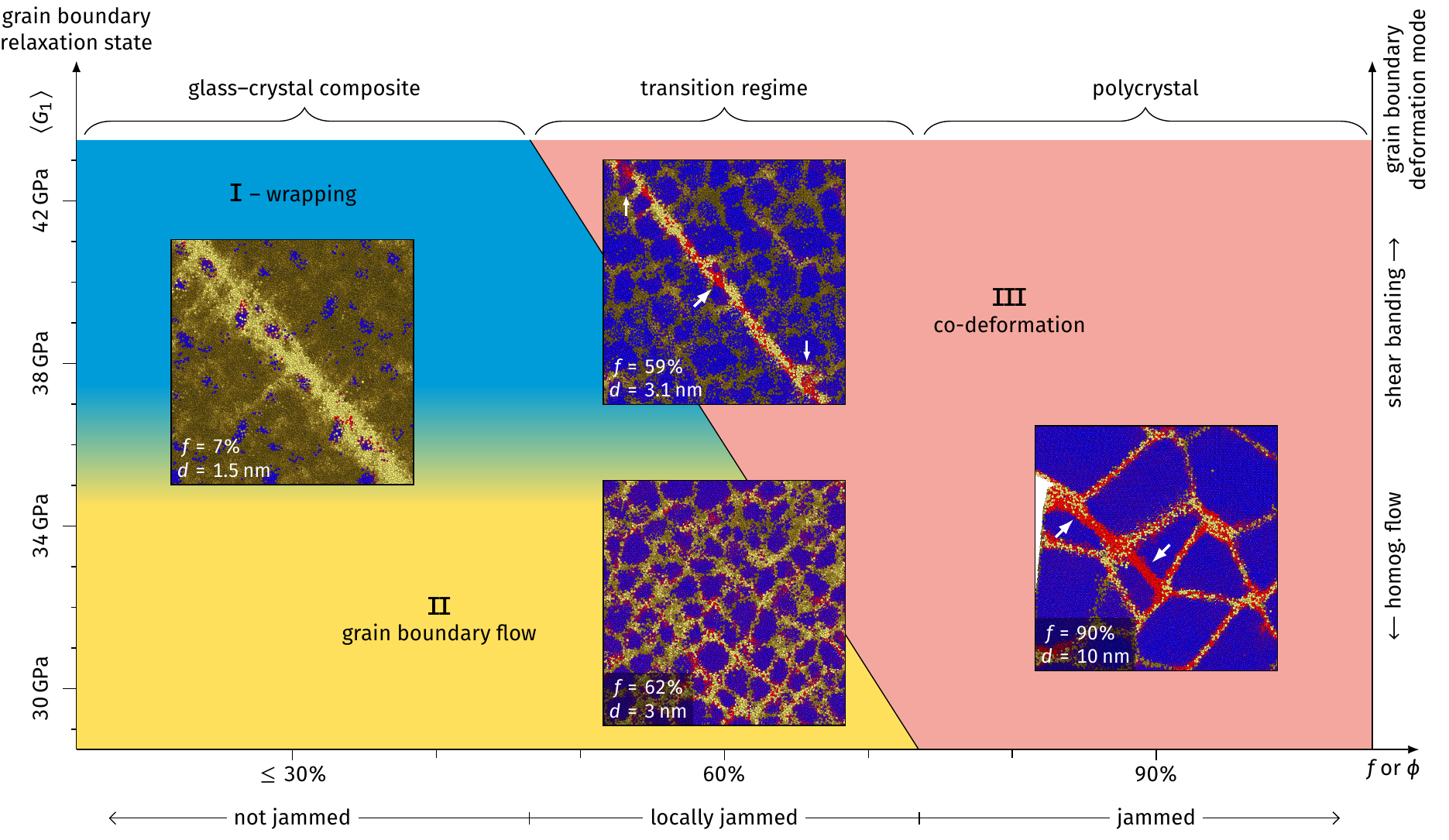}
  \caption{Mechanism map of the crossover from glassy to crystalline
    plasticity as a jamming-like transition. For small crystalline
    volume fractions $f$, the plastic response depends only on the
    amorphous phase. Depending on its relaxation state (quantified by its
    local shear modulus $\langle G_1 \rangle$), we distinguish
    two regimes: (I) The formation of
    a single shear band that can avoid the crystallites by
    ``wrapping'' around them and (II) a homogeneous flow of the glassy
    phase. The distinction is solely controlled by the relaxation
    state of the amorphous phase, be it bulk glass or grain
    boundary. With increasing crystalline volume fraction $f$ (and
    thereby increasing packing density $\phi$ of the crystallites),
    clusters of crystallites become locally jammed. While the
    homogeneously flowing grain boundary can deform while avoiding
    these areas, the localised shear band can only propagate by
    cutting through the crystallites. Therefore, the onset of regime
    III, i.e., the co-deformation of both phases, also depends on the state
    of the grain boundary.  Finally, a further increase of $f$ leads to
    full jamming and co-deformation independent of the grain boundary
    state. This is the usual scenario in nanocrystals with grain sizes
    above roughly \SI{10}{nm} and in coarse-grained polycrystals.}
  \label{fig:mechanism-map}
\end{figure*}

\subsection{Change of mechanisms as a jamming transition}

While shear localisation does not strongly influence the strength of
the glass \cite{Albe2013}, it nevertheless appears to be important for
the difference of mechanisms between sample XI and the 3\,nm Voronoi
construction, as evidenced by Figs.~\ref{fig:yield-over-shearmod},
\ref{fig:shear-localization-annealed}, and
\ref{fig:shear-localization-voronoi}.  With increasing $f$, the
contribution of the yield strength of the crystallites increases,
finally exceeding the strength of the grain boundary.  As
illustrated in Fig.~\ref{fig:mechanism-map}, this behaviour can be
understood as a kind of jamming transition, which is known from
granular matter \cite{Cates1998, Liu2010, Torquato2010}: At lower
packing density $\phi$, the granular particles can flow freely. In our
analogy, this means that the shear band can wrap around crystallites
and that homogeneous flow around the crystallites is possible. When
increasing $\phi$---which is roughly equivalent to $f$ in our case---we locally
get a few clusters of jammed crystallites. In the case of localised
deformation, the shear band samples only a small volume. This volume
itself is jammed and thus the crystallites either need to be cut, or a
new shear band nucleated. The latter case probably costs more energy
and there is already a stress concentration at the shear band
front. This does not concern the homogeneously flowing grain boundary,
though, because a local jam in one area can be avoided by flowing
somewhere else.  In terms of jamming, this corresponds to keeping the
number of contacts constant, but increasing the system size, thereby
unjamming the system \cite{Moukarzel2012, Goodrich2012}. The final
state is a completely jammed system, which corresponds to the
classical picture of a polycrystal:  Here, the crystallites always carry
a large part of the plasticity (i.e., they need to be
``cut'') independent of the grain boundary state. In our system, no Hall--Petch
effect can be observed due to the brittleness of the Laves phases and the yield stress
reaches a maximum.

The classical picture of the jamming transition corresponds more
closely to completely homogeneous flow of the grain boundary. For hard
spheres, a critical packing density of around $\phi = 64\%$ is often
found \cite{Liu2010, Torquato2010}. Looking at
Fig.~\ref{fig:yield-over-shearmod}, we find that this roughly
corresponds to the critical $f$ for cutting crystallites, which lies
somewhere between the \SI{3}{nm} and \SI{5}{nm} samples. This is
remarkable, since the assumptions of the hard sphere model barely hold
in our system. Nonetheless, the model appears to apply at least
approximately. We can also see that the different yield stresses of
the C14 and the C15 phase only come into play for large $f$; before
that point the strength of the sample is still grain-boundary dominated. We note
that the volume fraction $f$ is independent of any length scale. The
length scale gets reintroduced by the thickness of the grain boundaries,
which controls $f$ for densely packed crystallites of a given
diameter. For typical grain boundary widths in the nanometre range,
the jamming transition occurs somewhere below a grain size of
\SI{10}{nm}.

The possibility of purely grain-boundary-mediated deformation has been
reported before for Ni \cite{Rupert2013a}, where it is enabled by a
free percolation path for shear bands through the sample.  Our
results furthermore indicate that the participation of the crystalline
phase in the deformation also depends on the intrinsic proclivity
towards strain localisation in the amorphous phase.  Pure metals in
simulations at nanometre-scale grain sizes always have an unrelaxed grain boundary,
since massive grain growth would set in on relaxation or annealing.  The softening
with decreasing grain sizes below \SI{10}{nm} that has been found
before \cite{Schiotz2003} is thus analogous to our Voronoi constructed
samples: For grain sizes increasing from \SI{3}{nm} to \SI{10}{nm}, the number
of crystallites participating in the deformation rises,
and with it the yield stress.  The only difference is that
the brittle failure of the crystallites is replaced
by plastic flow.  We confirmed the transferability of our observation to ductile
metals with example
simulations of nanocrystalline copper using the potential by Mishin
\textit{et al.}\ \cite{Mishin2001} (see Supplementary Fig.~S.7).

A maximum strength should be obtained when the grain boundary is very
relaxed and the deformation is localised, but when the grains are
small enough to suppress dislocation nucleation. Recent experimental
results on Mg-based nanocrystals in the form of a brittle Laves phase
with a large fraction of amorphous grain boundary confirm the
existence of such a regime \cite{Wu2017}: Shear-band-like deformation
occurs and cuts through the crystallites. The strength of the material
is very high compared to Mg-based polycrystals and glasses.

\subsection{Strengthening through grain boundary relaxation}

\begin{figure}[t]
  \centering
  \includegraphics[center]{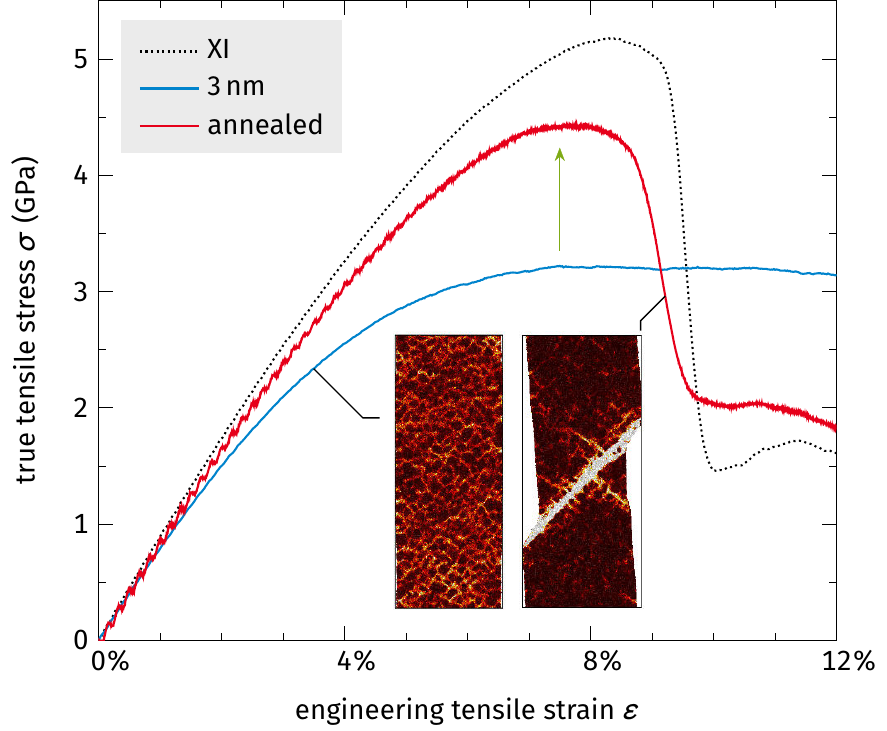}
  \caption{Effect of an annealing treatment on a Voronoi construction
    with C15 crystalline phase and \SI{3}{nm} grain size. While the
    untreated sample deforms homogeneously, the annealed sample
    localises the shear strain. This results in increased strength and
    brittleness. The stress--strain curve for sample XI, which has a
    comparable microstructure, is shown for comparison.}
  \label{fig:harden}
\end{figure}

Finally, since the mechanical properties of very fine-grained
nanocrystals depend largely on the properties of the amorphous phase,
it should be possible to modify the properties of the artificially
created sample with C15 phase and grain size \SI{3}{nm} by an annealing
procedure. We annealed the sample at \SI{900}{K} for \SI{4}{ns}. No
grain growth was observed.  Afterwards, we equilibrated the sample at
\SI{50}{K} and performed a tensile test with the same parameters as
before. Figure~\ref{fig:harden} shows that the structure now almost
resembles sample XI in its behaviour. The deformation is localised and
the strength is increased.

Repeating the procedure with the sample with grain size \SI{7}{nm}, we
find that the yield strength at \SI{50}{K} increases to around
\SI{5.1}{GPa}. Again, no grain growth was observable. This number is
comparable with the yield strength of sample XI and the sample with
\SI{10}{nm} grain size. Consequently, we conclude that an inverse
Hall--Petch effect is only found when the grain boundary is in a
rather unrelaxed state and deforms more homogeneously. In that case,
the strength of the material depends on the fraction of amorphous
boundary phase. Note that the samples I--X are less strong than XI,
but are not true nanocrystals due to the large amorphous volume
fraction.

Experiments show that annealing treatments of nanocrystalline metals
lead to grain boundary relaxation and strengthening, and for grain
sizes smaller than \SI{10}{nm} an increased strain localisation was
found \cite{Rupert2012a}. This is consistent with the current
observations and with a glass-like behaviour of the grain boundary. As
stated earlier, the strain localisation and strength in metallic
glasses depend on the relaxation state.  As such, the strengthening
observed in our simulations provides further evidence for STZ-like
mechanisms in the grain boundary.

\section{Conclusion}

In our MD simulations on \ce{Cu64Zr36} glass samples with different
volume fractions of brittle Laves phases, we find that the grain
boundary phase behaves like a metallic glass under constraint from the
abutting crystallites. The switch from glass-like to
grain-boundary-mediated plasticity depends on the crystalline volume
fraction and the grain boundary state. Tensile test simulations reveal
three regimes: (1) For low crystalline volume fractions, the system
behaves like a glass--crystal composite with small crystallites
\cite{Brink2016}, i.e., plastic flow is localised in the amorphous
phase. (2) With increasing crystalline volume fraction, clusters of
crystallites can become jammed. The behaviour of such a system depends
critically on the grain boundary relaxation state, which governs the
strain localisation tendencies in the amorphous phase: If the flow is
homogeneous, jammed clusters can be avoided and the deformation stays
confined to the grain boundary. If these clusters become larger, they
nevertheless contribute partially to the material strength, as would
be expected from a simple composite model. If the flow is instead
localised in a shear band, the jammed clusters present obstacles to
the deformation and need to be cut. Thus, they contribute fully to the
strength of the material. (3) At more typical experimental grain sizes
$\geq \SI{10}{nm}$, the system is jammed completely and the grain
boundary can no longer be the sole carrier of plasticity, leading to
co-deformation. In this typical polycrystalline regime, the
Hall--Petch scaling becomes valid.  These observations are
transferable to ductile metals, although the ``cutting'' of the
brittle crystallites is replaced by dislocation activity inside the
grains, and the transition regime is most likely less
sharp. Regarding the experimental observations of an inverse
Hall--Petch effect, we conclude that even polycrystals with very small
grain sizes only soften under specific conditions, namely when the
deformation is delocalised and grain-boundary dominated. Therefore,
experiments conducted at different grain sizes and with different
grain boundary compositions are not directly comparable.

\section*{Acknowledgements}
Financial support by the Deut\-sche For\-schungs\-ge\-mein\-schaft
(DFG) through project grant no.\ AL 578/6-2 is gratefully
acknowledged.
Calculations for this research were conducted on the Lichtenberg high
performance computer of the Technische Universit\"at Darmstadt.

\appendix
\renewcommand*{\thesection}{\appendixname\Alph{section}} %

\section{Change of glass transition temperature with annealing}
\label{sec:appendix:Tg}

\begin{figure}[b!]
  \centering
  \includegraphics[center]{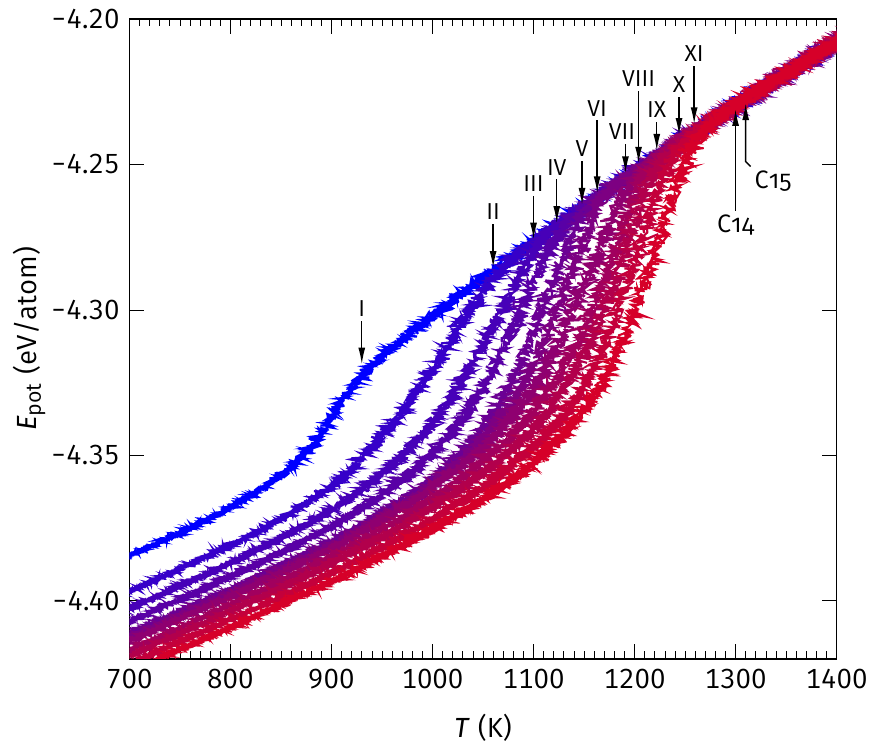}
  \caption{Change of glass transition temperature with annealing.
    Arrows indicate the upper bound for $T_g$ for the different
    samples, as well as an estimate for the melting points of the C14
    and C15 Laves phases.}
  \label{fig:Tg}
\end{figure}

We evaluated the change of glass transition temperature $T_g$ of the
samples I--XI by reheating them from \SI{50}{K} to \SI{1500}{K} with
$\dot{T} = \SI{0.1}{K/ps}$. The resulting potential energy over
temperature plot is shown in Fig.~\ref{fig:Tg}. An upper bound for the
glass transition temperature is indicated by arrows. Additionally, the
melting points of the C14/C15 Laves phases are indicated in the
plot. These were estimated by relaxing a liquid/crystal
interface at different temperatures. The transition temperature for
the samples continually rises with annealing time and no second
melting point for the contained crystallites can be observed. This
indicates that the liquefaction process for both phases occurs
simultaneously. The reason is most likely the small size of the
crystallites.

\section{Detection and analysis of the crystalline phase}
\label{sec:appendix:crystanal}

\begin{table}[b!]
  \centering
  \caption{Properties of the composites. For each sample, the
    crystalline volume fraction $f$, the average grain size $d$, the
    number density $\rho_n$, and the potential energy $E_\text{pot}$
    of the whole system at \SI{0}{K} is listed. Note that the average
    potential energy is partly lower than for the pure Laves phases
    (\SI{-4.496}{eV/atom} for C14 and \SI{-4.497}{eV/atom} for C15).
    This is due to the different composition (\ce{Cu64Zr36} vs.\
    \ce{Cu2Zr}), not due to the occurrence of a different phase.}
  \label{tab:properties}
  \newcolumntype{C}{>{\centering\arraybackslash}X}
  \begin{tabularx}{\linewidth}{CCCCc}
    \toprule
    Sample & $f$  & $d$  & $\rho_n$       & $E_\text{pot}$  \\
           & (\%) & (nm) & (\si{nm^{-3}}) & (\si{eV/atom}) \\
    \midrule
    I    & $\phantom{0}$0.0 &     & 63.15 & $-4.483$ \\
    II   & $\phantom{0}$6.9 & 1.5 & 63.27 & $-4.495$ \\
    III  &             14.4 & 1.5 & 63.33 & $-4.500$ \\
    IV   &             21.2 & 1.7 & 63.37 & $-4.505$ \\
    V    &             29.0 & 1.8 & 63.42 & $-4.509$ \\
    VI   &             35.9 & 1.9 & 63.43 & $-4.510$ \\
    VII  &             41.5 & 2.1 & 63.44 & $-4.512$ \\
    VIII &             47.2 & 2.3 & 63.47 & $-4.515$ \\
    IX   &             51.5 & 2.7 & 63.48 & $-4.517$ \\
    X    &             55.0 & 2.8 & 63.50 & $-4.519$ \\
    XI   &             59.1 & 3.1 & 63.53 & $-4.522$ \\
    \bottomrule
  \end{tabularx}
\end{table}

The software package \textsc{ovito} \cite{Stukowski2010} was used for
all analyses. Basic structural analysis was performed using Voronoi
tessellation, which divides the simulation cell into one polyhedron
around each atom \cite{Voronoi1908a, Brostow1998}. The polyhedra are
characterised by the Voronoi index
$\langle n_3, n_4, n_5, n_6 \rangle$, where $n_i$ denotes the number
of $i$-edged faces of the polyhedron.
There exists no ready-made detection algorithm for Laves phases. In
lieu thereof, we used the following method: The zirconium atoms in the
Laves phases are characterised by a hexagonal (C14) or cubic (C15)
diamond superstructure \cite{DeGraef2007}.  Thus, they were identified
by applying a diamond structure identification algorithm
\cite{Maras2016}.  The copper atoms in both Laves phases appear as
Voronoi $\langle 0,0,12,0\rangle$ icosahedra \cite{DeGraef2007}. Thus,
those atoms that are either in a zirconium diamond superstructure or
that are copper icosahedra neighbouring such zirconium atoms, were
identified as belonging to a Laves phase.  We find that
crystallisation occurs almost immediately upon annealing, reaching a
saturation after roughly \SI{4}{\micro\second} (see
Fig.~\ref{fig:annealing-procedure}). The crystallites are a mix of C14
and C15 Laves phases, which have very similar cohesive energies in the
potential we used (\SI{-4.496}{eV/atom} for C14 and
\SI{-4.497}{eV/atom} for C15). The structural motifs of the
Laves phases---copper-centred icosahedra and Zr-centred
$\langle 0,0,12,4 \rangle$ (or ``Z16'') polyhedra
\cite{DeGraef2007}---also appear as low-energy configurations in the
glass \cite{Cheng2008b, Ding2014}. The grain boundary stays amorphous
in all samples.

In order to obtain an approximation for the grain sizes in the
annealed samples, we assumed that all crystallites are spherical and
equal in size.  The grain size $d$ is then
\begin{equation}
  d = \sqrt[3]{\frac{6 f N \Omega_\text{Laves}}{\pi n}},
\end{equation}
where $f$ is the crystalline volume fraction, $N$ the number of atoms
in the system, $\Omega_\text{Laves}$ the average atomic volume in the
Laves phase ($\approx \SI{15.31}{\angstrom^3}$), and $n$ the number of
crystallites obtained by cluster analysis. A detailed listing of the
samples' properties can be found in Table~\ref{tab:properties}.

\section{Nucleation of the crystallites}
\label{sec:appendix:nucleation}

\begin{figure}[]
  \centering
  \includegraphics[center]{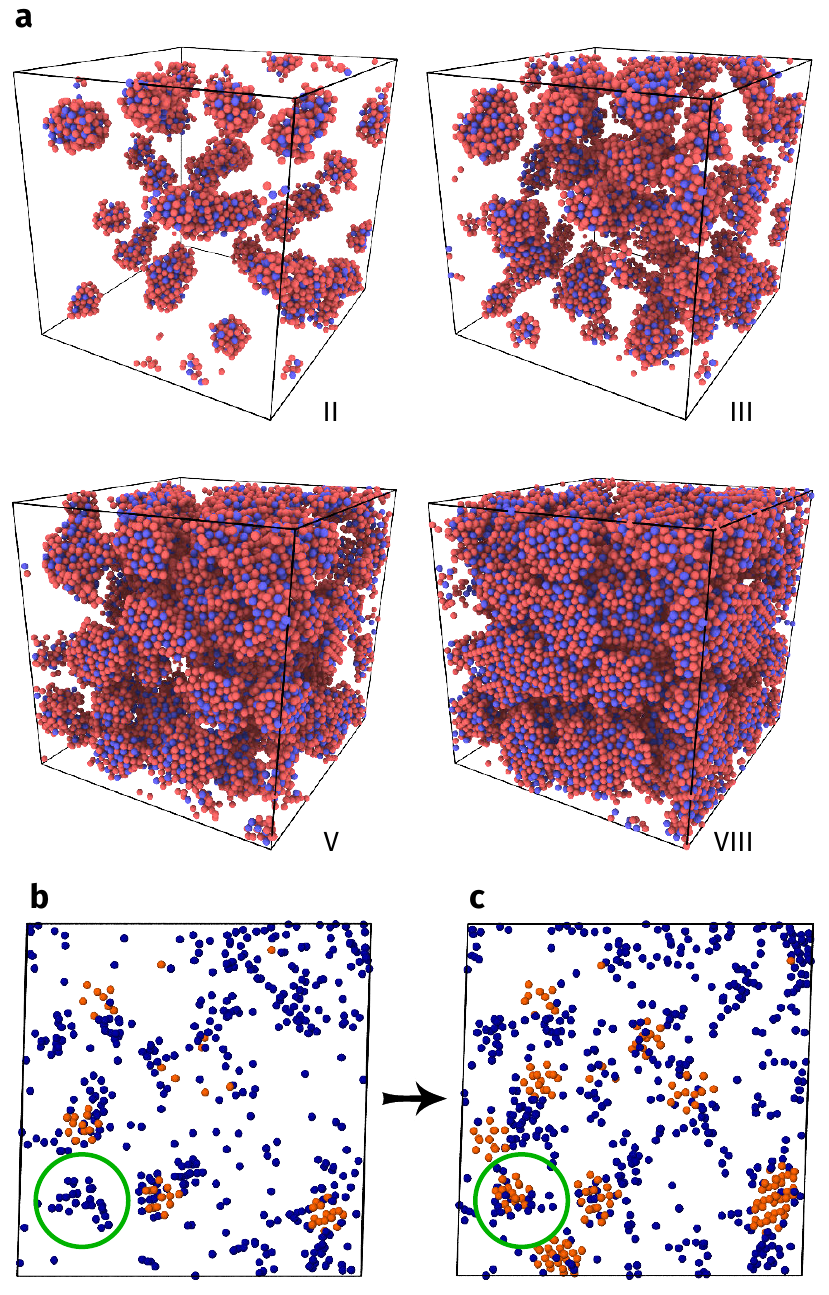}
  \caption{Snapshots of the crystallisation process. (a) Four
    exemplary snapshots of samples with different crystalline volume
    fraction. Only atoms identified as belonging to a Laves phase are
    rendered. (b)--(c) Slices through the sample after \SI{200}{ns}
    (b) and \SI{300}{ns} (c) annealing time are shown. The latter
    corresponds to sample II. All atoms except zirconium in Z16
    configuration were deleted for the visualisation. Orange atoms
    belong to a diamond superlattice, while blue atoms are located in
    amorphous regions.  The green circle indicates that
    crystallisation occurs preferentially in regions that already
    exhibit ordering tendencies.}
  \label{fig:annealing-snaps}
\end{figure}

Figure~\ref{fig:annealing-snaps} shows the crystallisation process.
We can see that the formation of the full crystalline phase is
preceded by the agglomeration of ordered clusters. This means that
crystallisation starts preferentially in the most ordered regions of
the glass. The reason is of course the structural similarity: Nelson
postulated some time ago that Frank--Kasper polyhedra, such as the
copper-centred icosahedra and the zirconium-centred Z16 polyhedra,
play an important structural role in metallic glasses and speculated
that the ideal glass may be a Frank--Kasper phase with infinite unit
cell \cite{Nelson1983, Nelson1983a}. Therefore, the Laves phases arise
most naturally in the most ordered regions of the glass.

\section{Analysis of the elastic properties}
\label{sec:appendix:elprop}

\begin{figure}[t!]
  \centering
  \includegraphics[center]{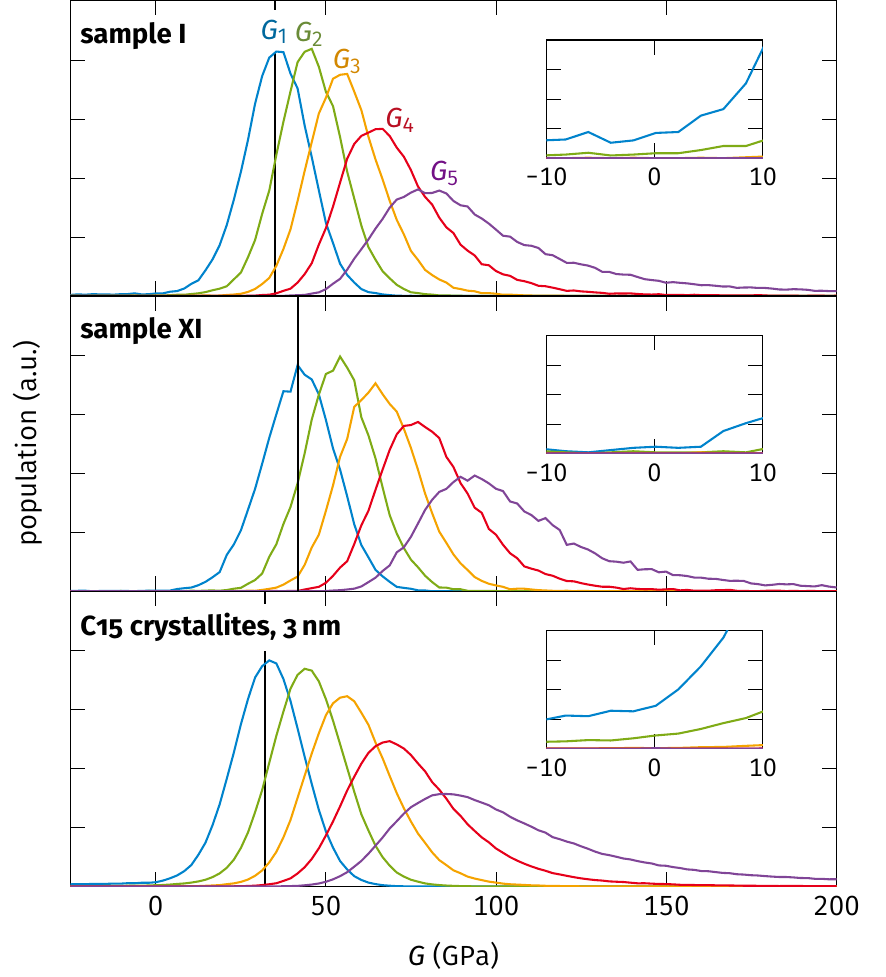}
  \caption{Histograms for the per-atom shear moduli $G_1$ to $G_5$ for
    selected samples. Only atoms in the amorphous phase are
    included. The vertical lines represent the median
    $\langle G_1 \rangle$. The insets show that negative shear moduli
    occur, especially for the less relaxed glasses. Scaling of the
    ordinate axis is the same for all three cases.}
  \label{fig:histograms-G}
\end{figure}

The per-atom shear moduli were calculated as described in
Ref.~\citenum{Brink2016a}: Different stress states were applied at
\SI{0}{K} and the stiffness tensor calculated with Hooke's law from
the strain response for every atom using \textsc{ovito}'s atomic
strain analysis \cite{Stukowski2010}. Diagonalisation of this tensor
yields five shear moduli \cite{Derlet2012}. Their distributions are
plotted exemplarily for three systems in
Fig.~\ref{fig:histograms-G}. The insets reveal that negative moduli
occur (as expected for amorphous systems, see
Ref.~\citenum{Derlet2012}), which decrease in number the more the
glass is relaxed. In order to extract a single number that most
closely relates to the mechanical strength of the phase, we decided on
an average of the lowest modulus, $G_1$. Because of large negative
outliers, the arithmetic mean is not representative (e.g., it would be
\SI{-32}{GPa} for sample I). Instead, we used the median.

We also note that the line in Fig.~\ref{fig:yield-over-shearmod}d
does not pass through the graph's origin as expected from experimental
data \cite{Johnson2005}. Essentially, $\langle G_1 \rangle$ does not
correspond to a macroscopic shear modulus. We know that a system
becomes mechanically unstable if a certain fraction of atoms has
negative moduli \cite{Schirmacher1998, Schirmacher2006, Schirmacher2008},
although a certain number can be accommodated by the surrounding matrix
\cite{Derlet2012}. Thus, $\langle G_1 \rangle$ is not necessarily zero
if the critical fraction is surpassed and the yield stress becomes
zero.

\section{Supplementary data}

\noindent
Supplementary data accompanies this article.

%

\clearpage


\section*{Supplementary material}

# From metallic glasses to nanocrystals: Molecular dynamics simulations on the crossover from glass-like to grain-boundary-mediated deformation behaviour

Tobias Brink and Karsten Albe

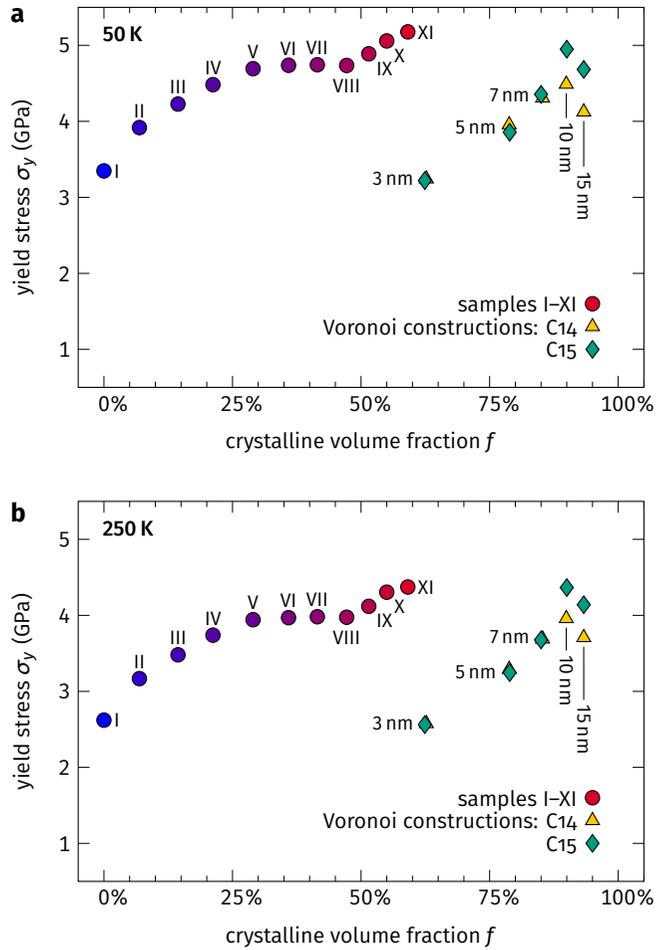

**Figure S.1:** Yield strength as a function of crystalline volume fraction for tensile tests at 50 K (a) and 250 K (b). The data points labelled I–XI refer to the samples with crystallites grown by annealing, while "Voronoi constructions" refers to artificially created polycrystals with different average grain sizes and crystal structures. Qualitatively, the behaviour is the same—independent of the temperature. Only the yield stress is uniformly reduced at 250 K.

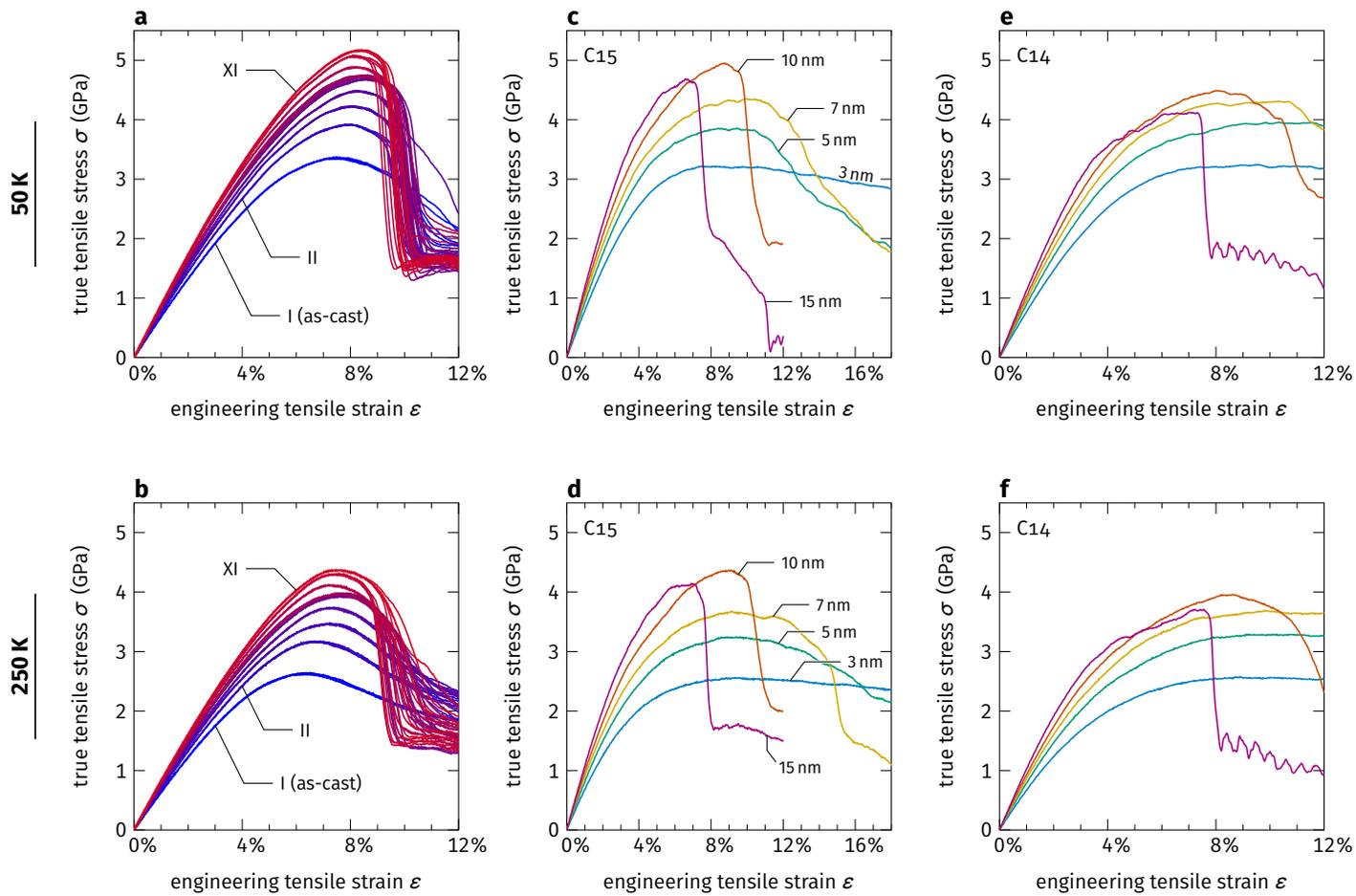

**Figure S.2:** Stress–strain curves of all samples.

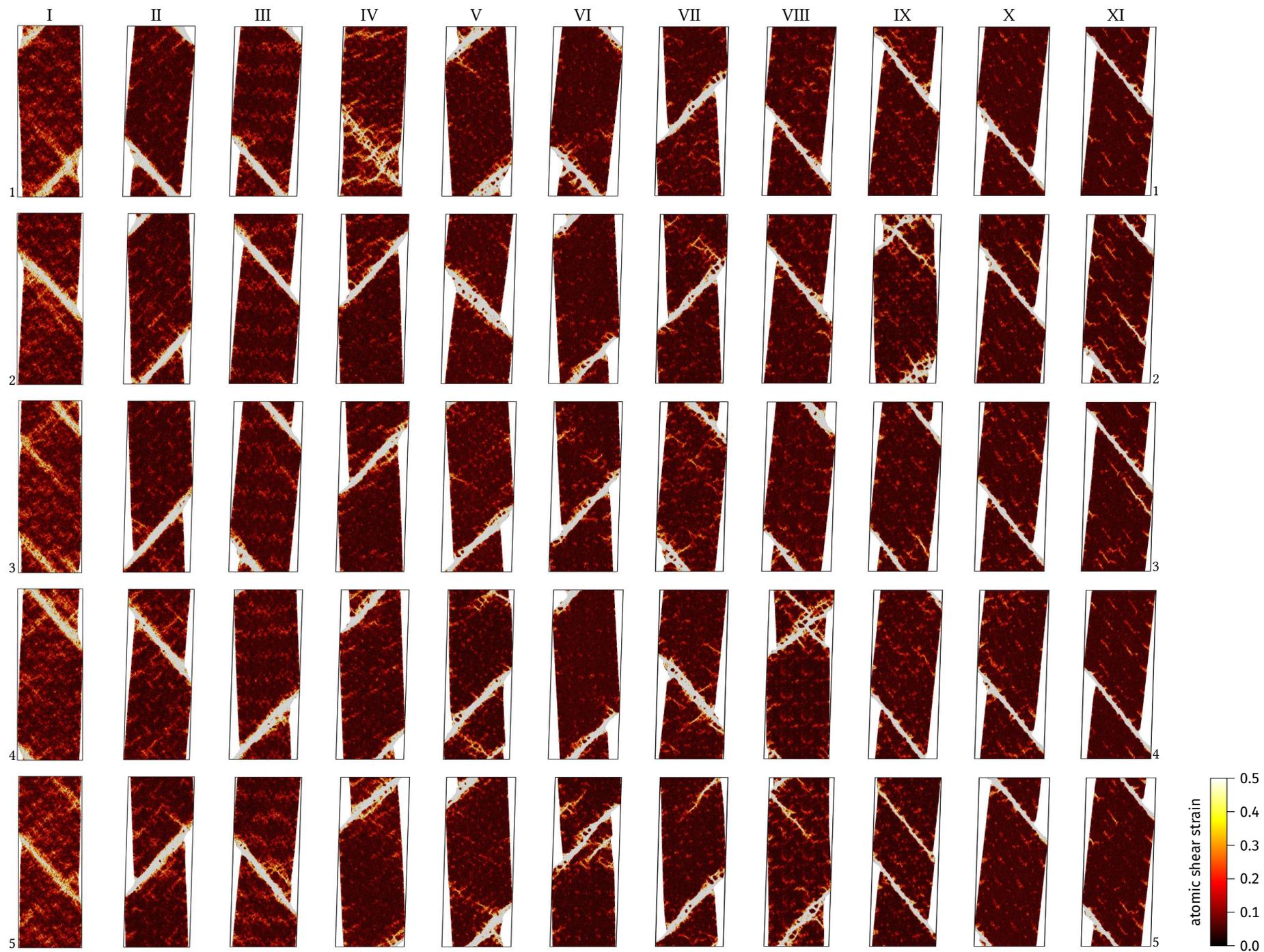

**Figure S.3:** Snapshots of samples after deformation at 50 K. Every row represents the results of a statistically independent run.

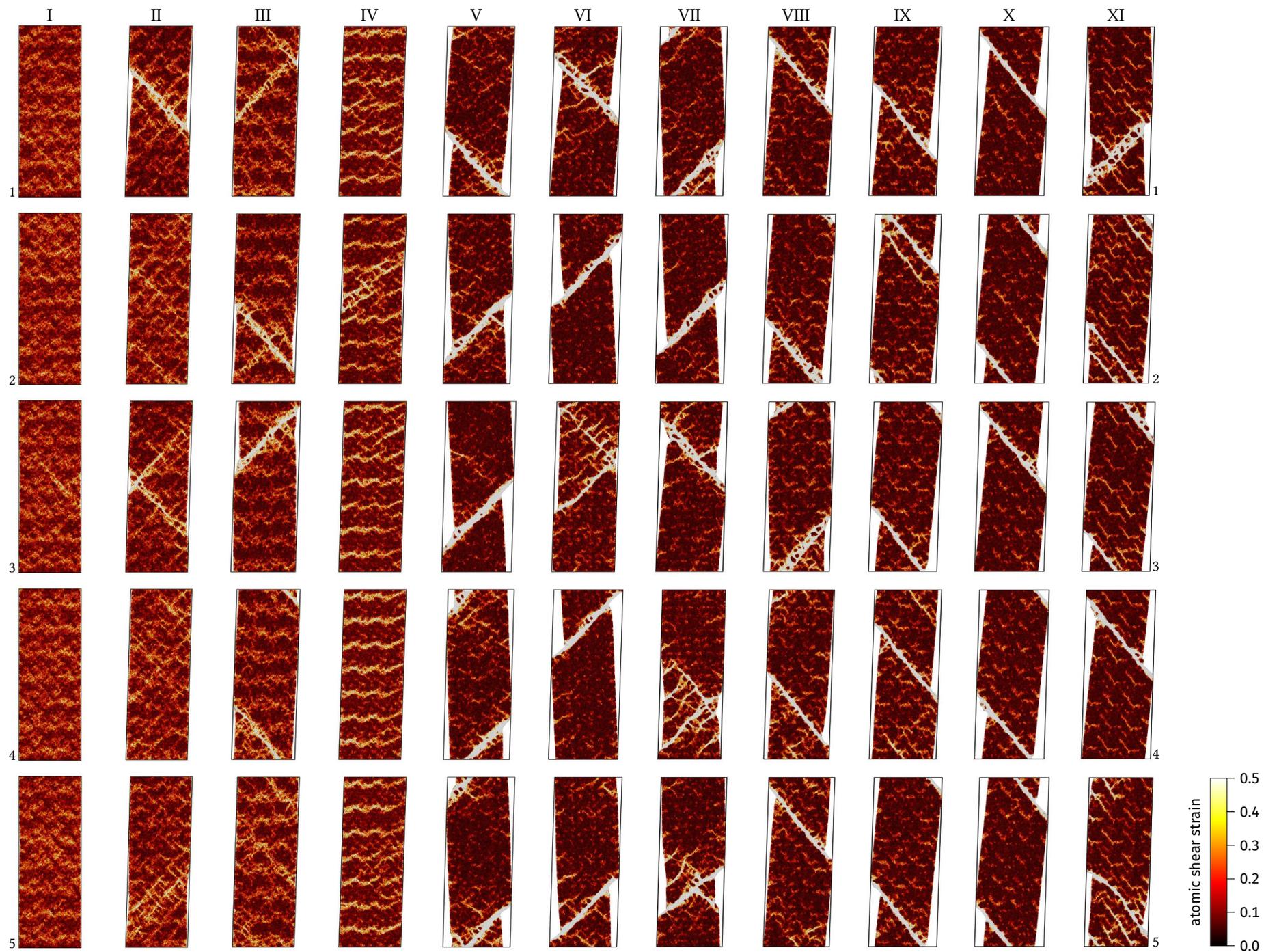

**Figure S.4:** Snapshots of samples after deformation at 250 K. Every row represents the results of a statistically independent run.

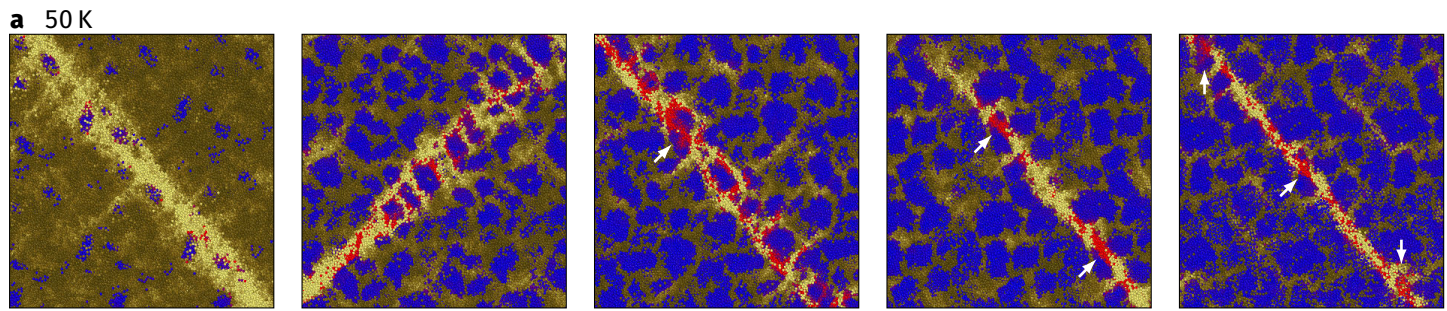
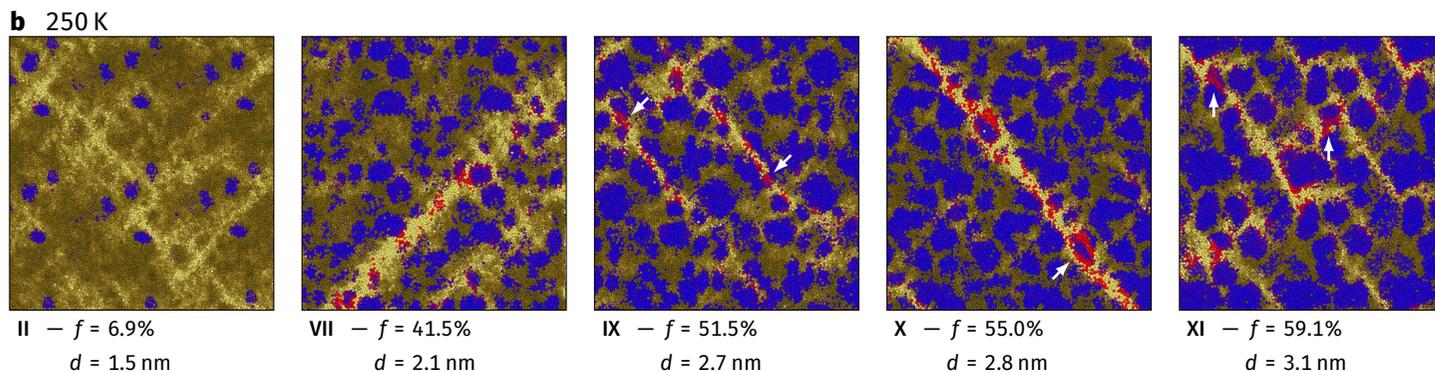
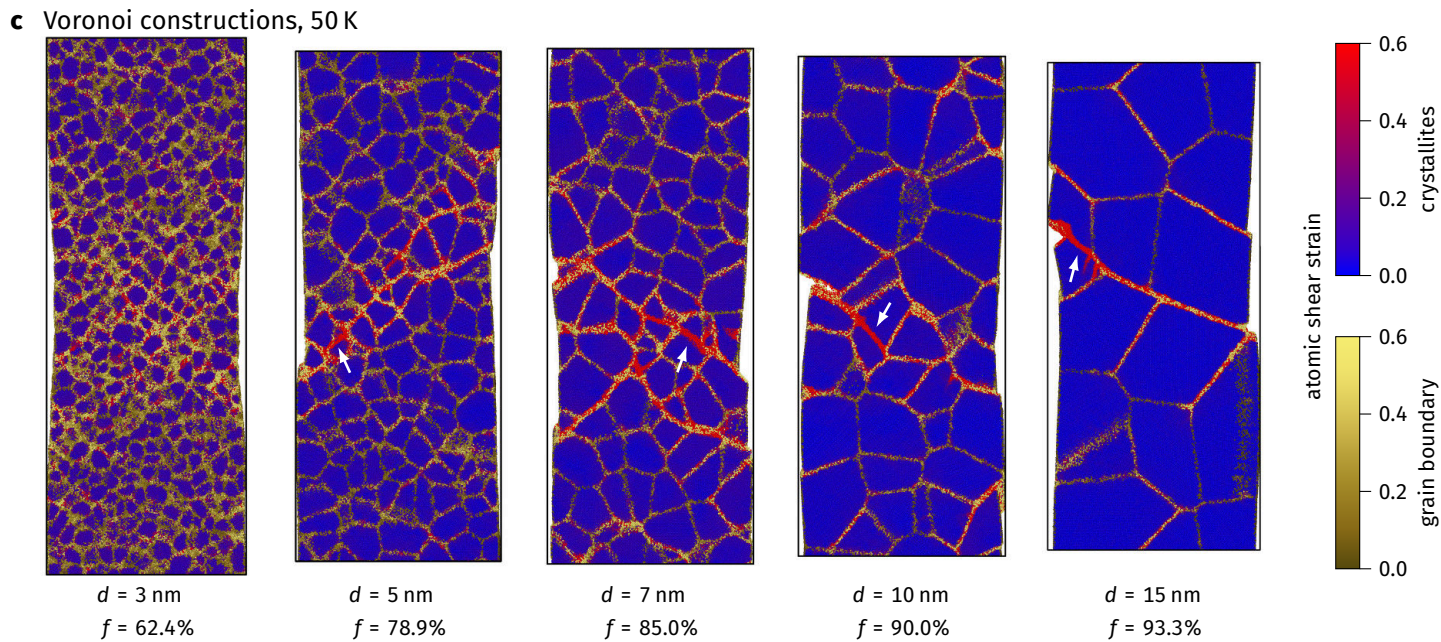

**Figure S.5:** Snapshots of selected samples after yield. The snapshots are coloured according to atomic strain, with different colour schemes for the crystallites and the amorphous phase. Arrows highlight were slip transfer from the amorphous phase into the crystallites takes place. (a) Detailed view of samples with crystallites grown by annealing, deformed at 50 K. Slip transfer and deformation of crystallites is only observed for large crystalline volume fractions. (b) The samples behave similar at 250 K, despite the reduced shear localisation. (c) The polycrystals by Voronoi construction usually exhibit localised deformation and slip transfer, except for the smallest grain size, at which a homogeneous grain-boundary-mediated flow occurs, which ultimately leads to necking. We only show samples with C15 crystal structure here, the C14 samples are qualitatively similar.

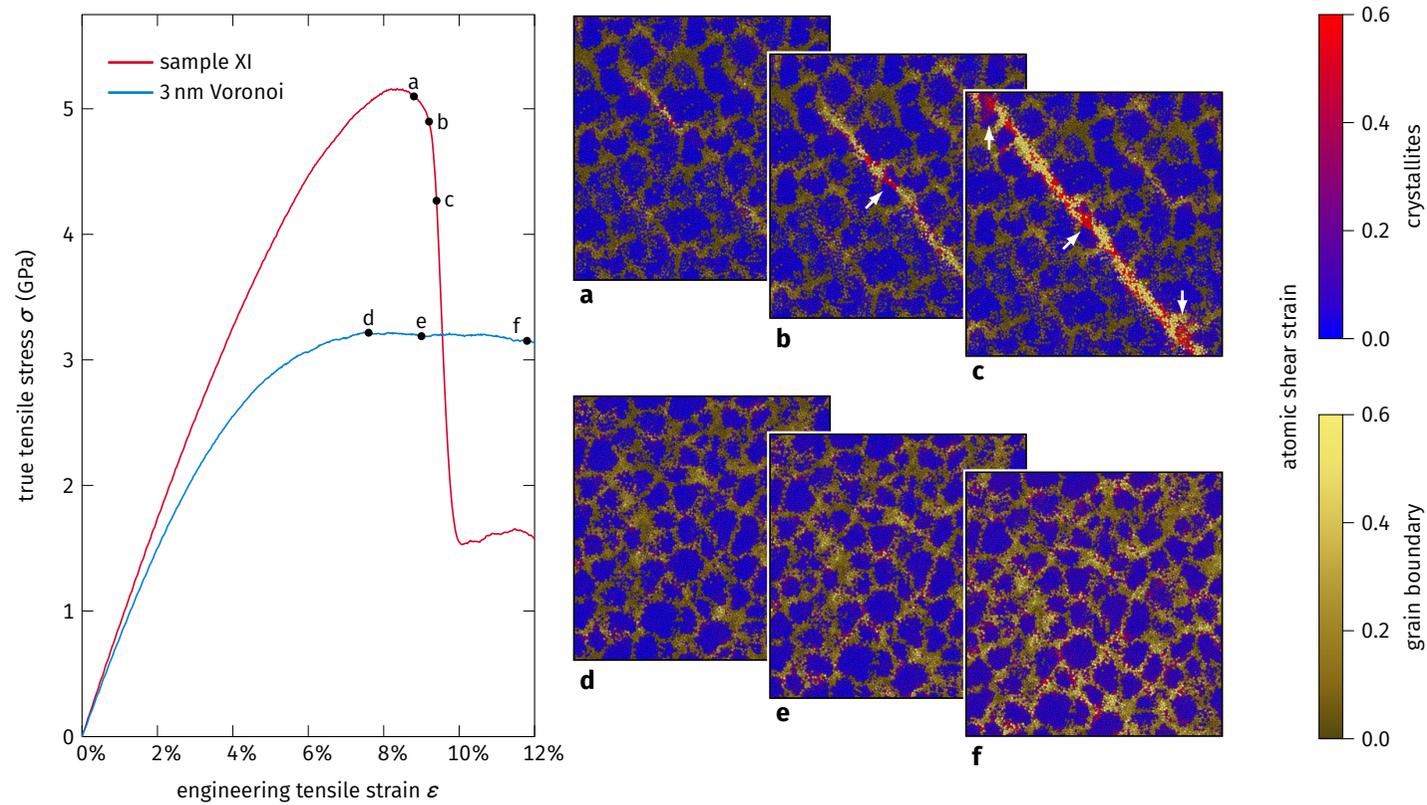

**Figure S.6:** Detailed yield process in sample XI and the Voronoi constructed sample with 3 nm C15 grains. Simulations performed at 50 K. (a) Yield in the relaxed sample starts in the amorphous phase, where it is confined by the surrounding crystallites. (b)–(c) The failure of the sample only occurs when the blocking crystallites are cut. (d)–(f) If the glass is less relaxed, though, the deformation is delocalised and purely grain-boundary-mediated flow can occur.

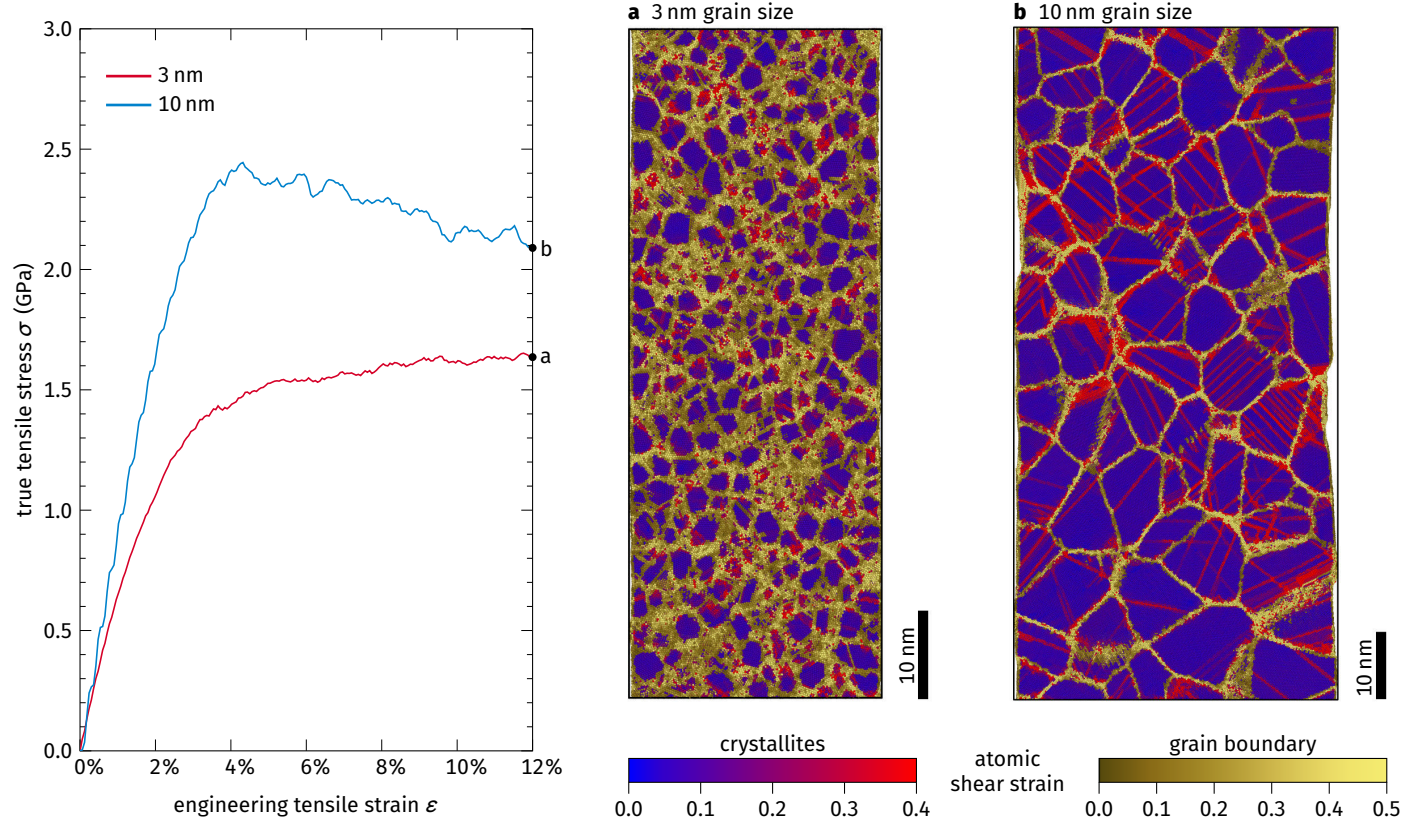

**Figure S.7:** Simulations of nanocrystalline copper with 3 nm and 10 nm grain sizes as an example of a ductile metal. The samples were created by Voronoi construction, minimised, equilibrated at 50 K for 1 ns, and finally deformed at $10^8$ s$^{-1}$. Deformation at 3 nm grain size is predominantly localised in the grain boundary, while the crystallites participate at 10 nm grain size. These results are qualitatively similar to the unrelaxed $Cu_{64}Zr_{36}$ samples. A relaxation of the grain boundary is not possible, since massive grain growth sets in immediately.


\end{document}